# Design and Commissioning of the first two CYRCé Extension Beamlines

E. Bouquerel[a], E. Traykov[a], C. Maazouzi[a], M. Rousseau[a], M. Pellicioli[a], J. Andrea[a],
T. Adam[a], P. Graehling[a], C. Mathieu[a], G. Heitz[a], M. Krauth[a], D. Oster[a],
T. Foehrenbacher[a], C. Ruescas[a], J. Schuler[a], U. Goerlach[a], C. Haas[a]

[a]*IPHC, UNISTRA, CNRS, 23 rue du Loess, 67200 Strasbourg, France*

**Abstract**

CYRCé is a TR24 cyclotron installed at the *Institut Pluridisciplinaire Hubert Curien* (IPHC) of Strasbourg operating at energies of 16-25 MeV and at intensities up to 400 $\mu$A. The accelerator is used to produce and provide radioelements for PET and for SPECT. In 2015, IPHC started to develop a platform with the aim of performing radiobiological experiments. The PRECy platform foresees to contain three-to-five experimental stations linked to beamlines expanded from the second exit port of the cyclotron. This extension allows devoting one of the beamlines for detector studies within the framework of the CMS project. The design, the development and the commissioning of the first two beamlines are discussed in this paper.

## 1. Introduction

In 2013, the *Institut Pluridisciplinaire Hubert Curien* (IPHC) of Strasbourg inaugurated its new accelerator manufactured by ACSI [1]. This cyclotron, baptized CYRCé (*Cyclotron pour la Recherche et l'Enseignement*), is a TR24 that operates at energy ranging from 16 to 25 MeV. Figure 1 shows the picture of CYRCé in his vault. With a very high current of up to 400 $\mu$A (upgradable to 1 mA) and proton energy up to 25 MeV, the cyclotron beams can produce commercial quantities of more than 20 isotopes. Table 1 shows the overall isotopes that can be available. $^{18}$F, $^{64}$Cu and $^{89}$Zr are the main radioelements produced by the facility.

Two years after the commissioning of CYRCé, IPHC started the development of a platform to perform radiobiological experiments using the proton beam exiting the cyclotron. This allows studying biological effects of energetic proton beams on molecules and cells, as well as on tumors implanted *in vivo* in small animal.

Email address: elian.bouquerel@iphc.cnrs.fr (E. Bouquerel)

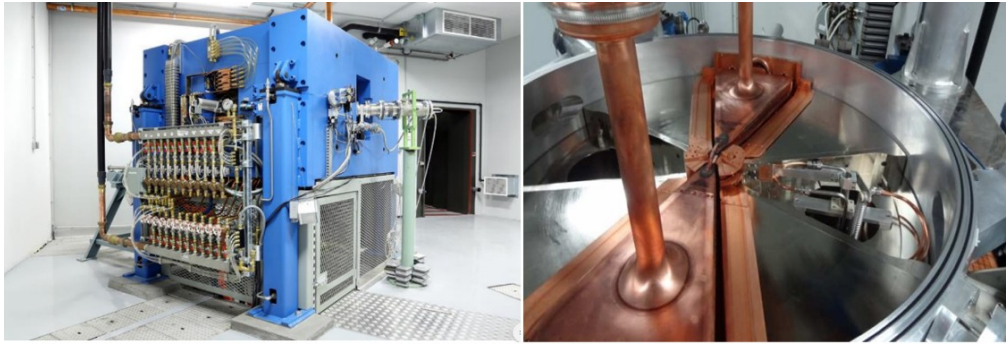
Figure 1: Pictures of CYRCé in the vault (left) and of the inside of the accelerator (right).

## 2. Goals and Scientific Contexts

### 2.1. *A platform for radiobiological experiments*

Radiotherapy is an essential therapeutic option in the life of patients with cancer in about 70% of cases. Presently radiation therapy uses x-rays or hadrons (mainly protons). The goals of modern radiotherapy are multiple: maximize the total dose in the tumor to increase damage to the tumor cells and to increase the lifespan of patients, minimize the dose deposition in healthy tissue to reduce side effects and preserve the quality of life of these patients. Protons use the physical characteristics of hadron radiation to reduce the dose in the critical organs. Unlike x-rays, the protons are charged particles allowing beam formation and manipulation for more precise targeting of the tumor. Indeed, the dose is delivered according to the Bragg peak beyond which the deposited dose is zero which is an advantage compared to heavier ion beams, such as carbon, which produce tails beyond the Bragg peak due to beam fragmentation reactions [2, 3].

The dose deposition is therefore mainly in the tumor and preserves the surrounding healthy tissue. In addition, the protons are biologically more effective than photons; it takes less of proton dose to obtain the same effect compared to the one obtained by photons. The relative biological effectiveness (RBE) allows estimating better the efficiency of a given radiation to produce a specific biological or pathological effect. The RBE is defined as the ratio of the radiation dose reference and the dose of the studied radiation, which produces the same biological effect. The results on the measurement of the RBE are highly variable. They depend on the studied cell, on the radiation energy and on the position of the Bragg peak during irradiation. Usually the RBE of protons is considered equal to 1 or 1.1. However, it can vary from 0.8 to more than 2 following these parameters and if the measurements are performed *in vitro* or *in vivo* [4].The nature of the effect of proton radiation in the cell is still very misunderstood and explorations should be pursued.

It is within this framework that the PRECy (Platform for Radiobiological Experiments from CYRCé) project proposes to undertake studies needed for a better understanding of the RBE *in vitro* and *in vivo* in small animals and the study of combination treatment.

A better understanding of these radiobiological effects may result in lowering the prescribed doses. The project consists in extracting and transporting the proton beam, accelerated by the cyclotron, out of the vault towards the experimental low energy radiobiology rooms for the *in vivo* study of the interaction of protons with the cells [5].

Table 1: Isotopes produced by CYRCé.

| Isotopes | Target/Reactio | $E_{protons}$ (MeV) | $\sigma_{max}$ (mbarn) |
|---|---|---|---|
| Actinium-225 | $^{226}$Ra(p,2n) | 15 | 700 |
| Arsenic-73 | $^{76}$Se(p,α) | 20 | 70 |
| Bromine-76 | $^{76}$Se(p,n) | 12 | 700 |
| Carbon-11 | $^{14}$N(p,α) | 7 | 250 |
| Cobalt-57 | $^{Nat}$Ni(p,x) | 25 | 600 |
| Copper-67 | $^{70}$Zn(p,α) | 15 | 15 |
| Copper-64 | $^{64}$Ni(p,n) | 11 | 700 |
| Fluor-18 | $^{18}$O(p,n) | 6 | 500 |
| Gallium-67 | $^{67}$Zn(p,n) | 10 | 600 |
| Gallium-68 | $^{68}$Zn(p,n) | 13 | 630 |
| Germanium-68 | $^{69}$Ga(p,2n) | 20 | 550 |
| Indium-111 | $^{111}$Pd(p,n) | 12 | 700 |
| Iodine-124 | $^{124}$Te(p,n) | 12 | 600 |
| Iodine-123 | $^{124}$Te(p,2n) | 24 | 1000 |
| Palladium-103 | $^{103}$Rh(p,n) | 10 | 500 |
| Rhenium-186 | W(p,n) | 10 | 17 |
| Sodium-22 | $^{22}$Ne(p,n) | 10 | 260 |
| Technetium-99m | $^{100}$Mo(p,2n) | 16 | 300 |
| Yttrium-86 | $^{86}$Sr(p,n) | 12 | 1000 |
| Zirconium-89 | $^{89}$Y(p,n) | 14 | 800 |

At 24 MeV, the maximum path of the protons in the material is 6 mm (water). By slowing down the beam, it is possible to cover a range of energy from a few 100 keV to 24 MeV and allow experimental measurements of the RBE on cell cultures and more fundamentally on the biomolecules. In other words, at low energy, it is possible to measure mainly the biological effects *in vitro* at the Bragg peak (at the level of the tumor) and *in vivo* in subcutaneous tumors implanted in small animals. The goal is to understand better the effects of the dose deposition at high linear energy transfer (LET) where biological effects are most important. A second experimental area is dedicated to CMS sensors qualification and industrial applications (radiation of materials, wear measurement, tests of electronic components …). Another alternative would consist in developing two beamlines, the first one dedicated to macro beams and a second for micro beams.

*2.2. A beamline for testing silicon modules*

The High Luminosity Large Hadron Collider (HL-LHC) is foreseen to start operating in 2026. It will allow an integrated luminosity of more than 3000 fb$^{-1}$ (inverse femtobarns) to be accumulated, ten times larger than during the present LHC phase. In order to resist to the high radiation level and to cope with a large number of superimposed events at each bunch crossing (up to 200), while keeping excellent performances, the Compact Muon Solenoid (CMS) detector needs to be thoroughly upgraded [6]. The CMS group at IPHC, strongly involved in the development of the new tracker, contributes to the quality tests of the new silicon modules: continuous development of the firmware of the acquisition system using a bench test at the laboratory; responsibility for data acquisition in beam test; analysis of the beam test data and characterization of the new silicon sensors. The upgraded tracker should have both higher granularity, better radiation handling, less material budget and higher geometrical acceptance. The construction of the radiobiology platform (PRECy) gives a great opportunity to develop, in parallel, a beamline specifically dedicated to the tests of silicon modules and its data acquisition, useful during the long LHC shutdown. Indeed, the conditions of the beam delivered by CYRCé allows performing high rate tests with high-energy deposits at a frequency close to that of the LHC. Long-term irradiation tests are also in consideration.

## 3. Beam Dynamics Investigations

### 3.1. Beam Requirements

As mentioned in the previous section, the main goal of the first PRECy beamline is to measure the biological effects at the level of cells. To achieve these measurements, sets of wells enclosing cells, are placed at the experimental station. Each set contains from 12 to 96 wells. The irradiation of each well is performed with a homogeneous beam transverse profile considering a constant deposition in depth [5]. A controlled bench moves the wells in the transverse directions in order for them to be hit by the protons. The energy and the intensity of the protons ranges from 24 MeV to few keV and from 100 pA to 10 nA respectively. The samples receive doses ranging from 0.01 to 6 kGy/min (up to 10 Gy in 100ms for FLASH radiations).

The experimental station located at the end of the other beamline is dedicated to the studies of CMS detectors. The set up consists of pixels and of two types of modules that are used in the Outer Tracker of the CMS detector (this set up is not presented in this article). These modules, called 2S and PS, are made of a stack of two silicon sensors with gap of 1.6 to 4.0 mm respectively [7]. The irradiation should be performed at high rates and high occupancies on full size of the samples: 40 MHz (6.4 pA), 6.25 MHz (1 pA) and 6.25 kHz (1 fA). The maximal beam spot needed is 30 mm diameter. Slow irradiation of detectors is possible with beams of up to 100 nA. At 25 MeV beam energy, the minimum ionization is foreseen to be at least 11 times greater than the one used for the LHC experiment.

### 3.2. Configuration of the experimental area

CYRCé has been installed into a vault of dimensions 6.5 m x 7.5 m. The distance between the second exit port of the cyclotron and the wall is 6 m (including its thickness). The experimental area is 15 m long and 13 m large and is separated from the vault by a 2 m thick concrete wall. Fig.2 shows the schematic layout of the overall area.

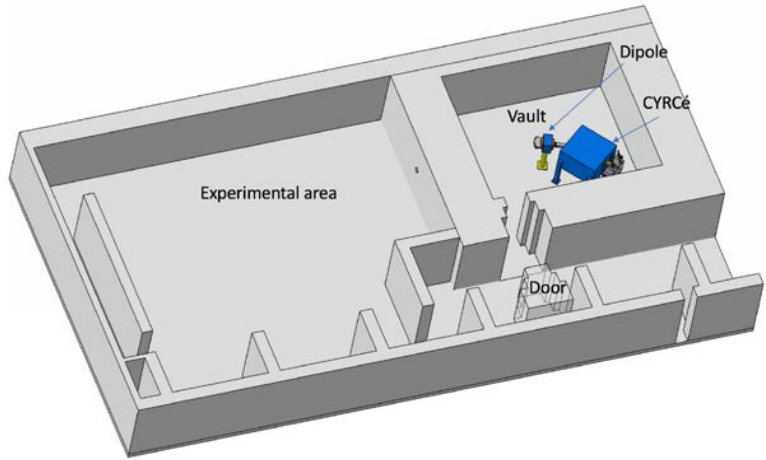

Figure 2: Schematic layout of the CYRCé location. A dipole, manufactured by ACSI [1], is connected to the second exit port of the cyclotron.

### 3.3. Beam parameters from the TR24

To design and define the optical elements mandatory to provide an efficient beam through the future beamlines, the proton beam delivered from the cyclotron has to be clearly characterized. Table 2 presents the physical parameters of the CYRCé cyclotron and the exiting proton beam.

Among the most important parameters necessary for designing a beamline are the transverse emittances. Indeed, the emittances give information on the size and the divergence of a beam. An emittance is defined as the six-dimensional distribution of all position coordinates along the three configuration-space directions and the associated velocity coordinates. Typically, it is projected into the two-dimensional subsets (x, x'), (y, y') and (z, z'). Several methods were used to estimate the values of the transverse emittances of the proton beam extracted from the TR24 cyclotron [8, 9]. The investigations to estimate these values allowed giving ranges and upper limits of these parameters (Table 3).

Due to many settings related to the cyclotron influencing the beam emittance, differences in the estimations were observed according to the measurement and the method used. For some cyclotron extraction tuning, the horizontal beam profiles had asymmetric structures (double peaks) and larger emittance values suggesting a strong dependence of the horizontal emittance on the tuning parameters.

Table 2: CYRCé characteristics.

| Parameters | Values |
| --- | --- |
| Charge | -1 |
| Number of dee | 2 |
| Max. extraction radius (cm) | 51 |
| Injection energy (keV) | 30 |
| Max. current ($\mu$A) | 400 |
| Extraction energy (MeV) | 16 to 25.4 |
| Momentum (MeV/c) | 218.033 |
| $\gamma$-1 | 0.026 |
| $\beta$ | 0.226 |
| $\beta\rho$ (T.m) | 0.727 |
| Time structure | CW (85 MHz RF) |
| Beam profile | Gaussian |

Table 3: RMS values of the transverse emittances (in mm.mrad) according to the method used and the beam energy (in MeV).

| Method | $x$ | $\Delta x$ | $y$ | $\Delta y$ | Energy |
| --- | --- | --- | --- | --- | --- |
| Quad scan | 1.90 | 1.30 | 3.70 | 1.40 | 25 |
| Slit and grid | 1.40 | 0.20 | 5.40 | 0.20 | 25 |
| Multiple profilers | 3.03 | 0.22 | 5.96 | 0.16 | 25 |
| Multiple profilers | 5.81 | 0.48 | 6.94 | 0.19 | 18 |

Nevertheless, results agreed that the vertical transverse emittance was between 1.2 and 3.9 times larger than the horizontal one at proton energy of 25 MeV. In order to consider the worst case and to avoid transmission losses, the upper limits of the transverse emittances were used as input parameters when designing the beamlines.

### 3.3.1. Multi-structures of the extracted beam

Multiple peak structures were observed in the horizontal plane when performing beam profile measurements to estimate values of the emittances (Fig.3).

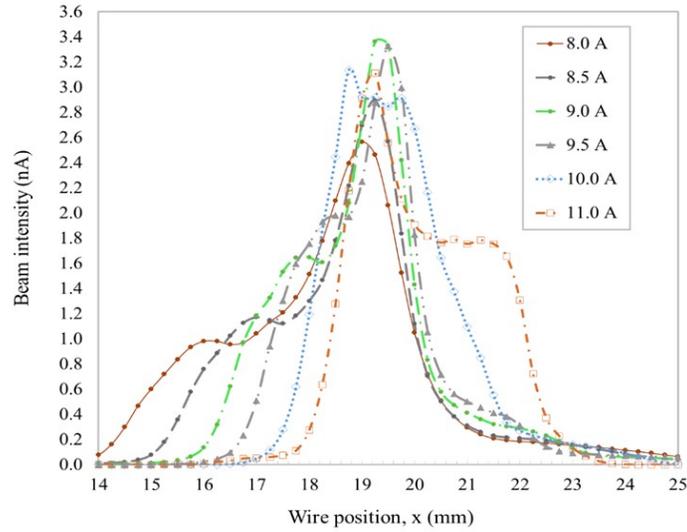

Figure 3: Beam profile in the horizontal plane when varying the current of one of the quadrupoles (QB) located at the extraction of the TR24 cyclotron (color should be used).

These structures in the profiles are known phenomena when extracting a beam from a cyclotron. Indeed, the reason is that the beam from two (or more) orbits with different energies are extracted at the same time. This is related to the foil position and the extraction angles as shown schematically in Fig.4.

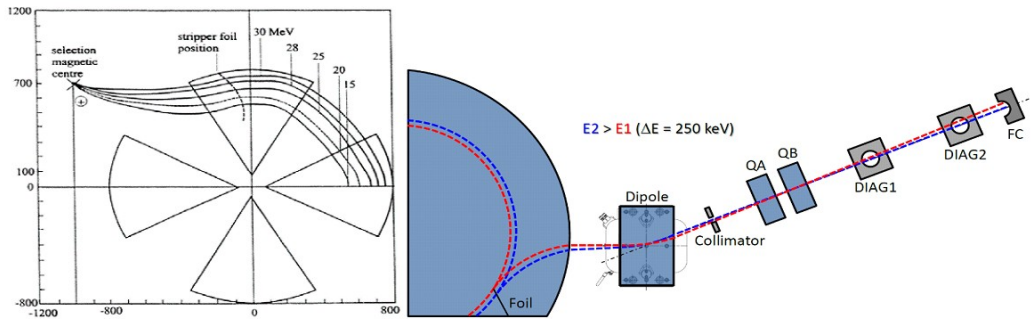

Figure 4: Particle trajectories inside a cyclotron (left) and during the extraction process (right) according to its kinetic energy. QA and QB refer to the quadrupoles. DIAG1 and DIAG2 are diagnostics. FC refers to Faraday cup.

To illustrate the effects observed during the experiment, different beams propagating within a beamline were simulated using TraceWin [10] and combined (Fig.5). As for the experiment to determine the values of the transverse emittances [8], this beamline is implemented by the use of a dipole, two quadrupoles and various diagnostics (slits, profilers and Faraday cups).
The combination of the different beams arriving into the virtual diagnostics confirmed the formation of a multiple peak structure in the beam profile in the horizontal plane (Fig.6). The observed structures were verified and also explained in detail by dedicated cyclotron extraction simulations, which were performed as part of an IN2P3-JINR collaboration [11].

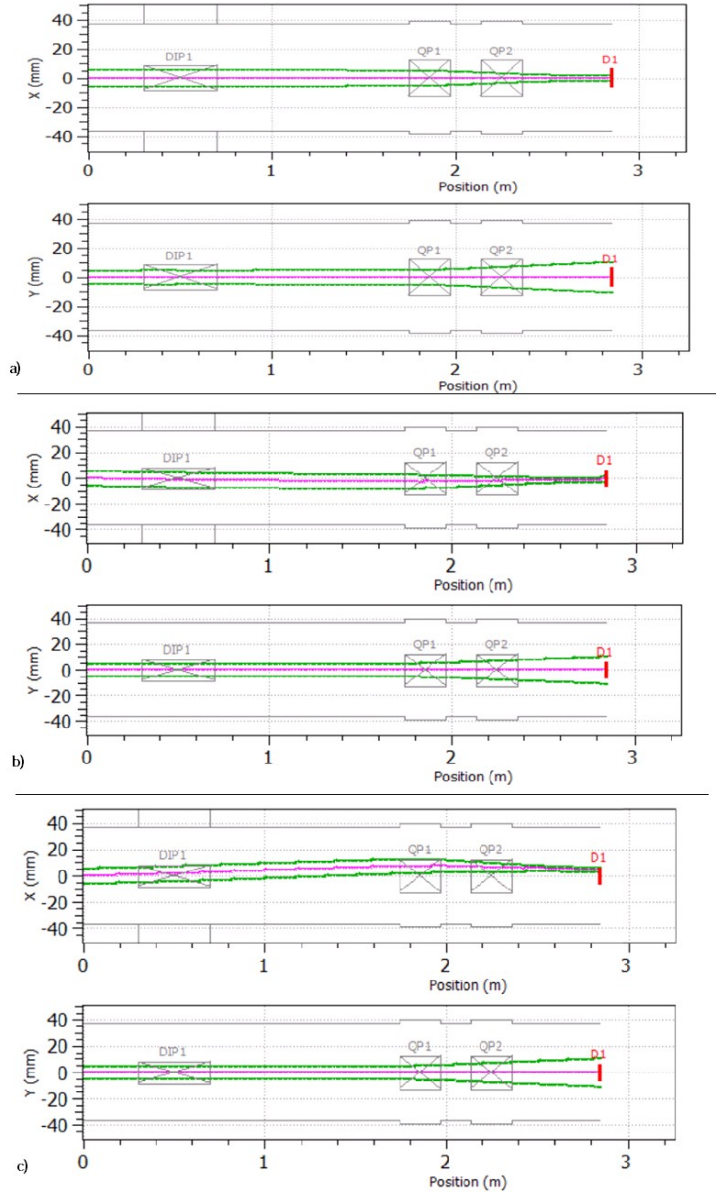

Figure 5: RMS transverse beam envelopes through a beamline composed of a dipole (DIP1), two quadrupoles (QP1 and QP2) and a diagnostic (D1). The kinetic energy of the protons is; case 1: 25 MeV (a); case 2: 24.75 MeV / -3 mrad off axis (b) and case 3: 25.25 MeV / +3 mrad off axis (c).

*3.4.* Design of the Cyclotron Beam Transfer Line (CBTL)

The transport line from the cyclotron to the experimental area has three distinct objectives: deliver a beam that is well centered, well focused and with a good transmission coefficient (minimal losses). A five-exit dipole is located after the vault wall allowing directing the beam either towards the PRECy or towards the CMS beamlines with the three remaining exits allowing the implementation of more user beamlines in the future. One of the two exit ports of the cyclotron is connected to a dipole magnet. This dipole, manufactured by ACSI [1], allows switching the extracted beam with a deflection angle of ±22°,to either the direction of the experimental area or to a short beamline specially dedicated to quick experiments to take place into the vault while the conception of the main beamline. Typical operating values are 0.75 T for the field and around 120 A for the current. Fig.7 shows the dipole and Table 4 lists its main operating parameters. The distance between the exit of the dipole switcher and the wall of the vault is 3.15 m.

The parameters used to design the CBTL are given in Table 5. As mentioned previously, the upper limits of the transverse emittances estimated in [8] were taken to perform the simulations.

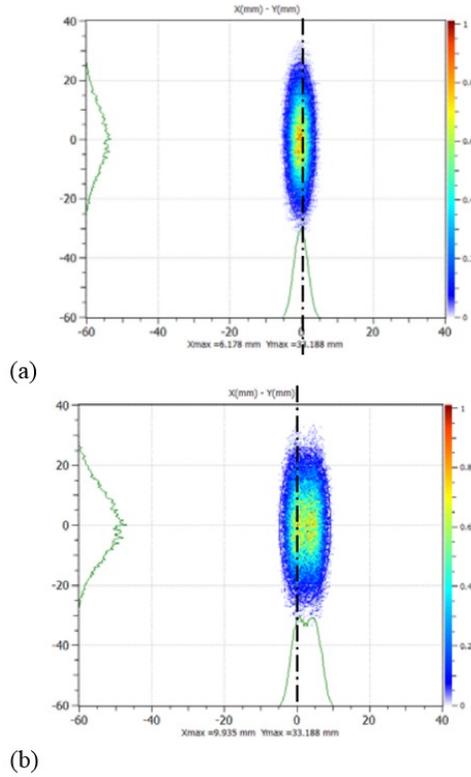

(a)

(b)

Figure 6: Particle distribution when case 1 is combined with case 2 (a) and when case 2 is combined with case 3 (b).

Table 4: Dipole main parameters [1]

| Parameters | Val |
| --- | --- |
| Max. Field (T) | 0.8 |
| Effective Length (mm) | 400 |
| Actual Length (mm) | 350 |
| Pole Gap (mm) | 50 |
| Current (A) | 130 |
| Voltage (V) | 35 |
| Cooling (L/min) | ≥15 |
| Total Weight (kg) | 380 |

Table 5: Beam parameters according to the energy of the protons [1]

| Parameters | Values at 25 MeV | Values at 18 MeV |
|---|---|---|
| Particle | proton | proton |
| $\gamma$ | 1.0256 | - |
| $\beta$ | 0.220 | - |
| Magnetic rigidity, $\beta\rho$ (T.m) | 0.727 | 0.615 |
| Emittances, $x^{rms}$, $y^{rms}$ ($\pi$ mm mrad) | 3.03 - 5.93 | 5.81 - 6.94 |
| Dp/p (%) | 1 | 1 |

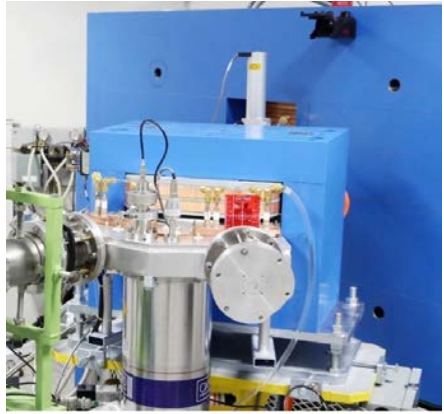

Figure 7: Picture of the dipole combo switcher at the exit of CYRCé.

One of the main conditions set up while doing the design and the optimization of the beamline and its optical elements is having a beam waist at the entrance of the five-exit dipole located after the wall connecting the vault to the experimental area. Investigations showed that beam sizes of less than 20 mm is achieved after the wall using a doublet of quadrupoles similar to the ones used for the estimation of the transverse emittances. Indeed, gradients of less than 7.3 T/m are sufficient to achieve the required beam size (Fig.8).

Simulations allowed defining the main parameters for the quadrupoles including their precise locations. Table 6 shows the characteristics of the doublet manufactured by SigmaPhi [12].

The extraction process of the beam from the cyclotron can induce off axis beam propagations. Therefore, a steerer located at the exit of the dipole is necessary to ensure the good alignment of the beam before it reaches the optical devices. Collimators and moveable slits are used to shape the beam before it enters the focusing system and the experimental area. Beam diagnostics such as profilers are used for the alignment of the beam. Faraday cups measure the beam current and can stop the beam in case of emergency. Fig.9 shows the 3D view of the CBTL with the implemented equipment.

The five exits of the dipole located after the wall offer the possibility to deflect the beam to 0°, ±22° and ±40° (Fig.10). The -22° angle exit is dedicated to the PRECy beamline and the +22° for CMS.

Table 7 shows the characteristics of the dipole, manufactured by SigmaPhi. The horizontal good field region is estimated ±34 mm.

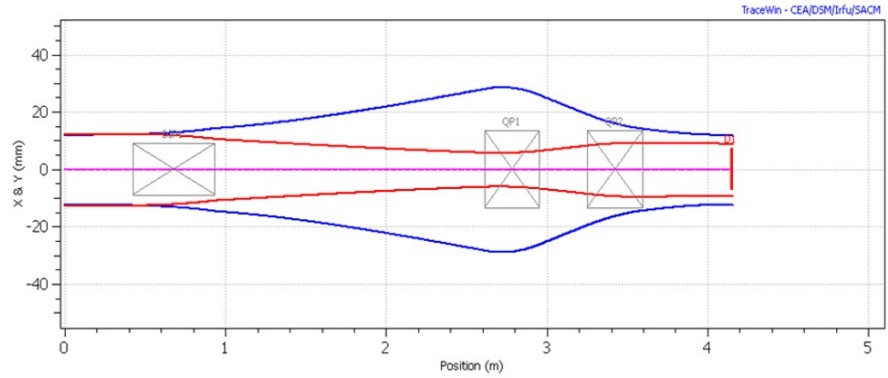

Figure 8: Beam envelopes (4σ) from the exit of the cyclotron to the entrance of the dipole in the experimental area. DIP1, QP1, QP2 and D1 stands for Dipole, quadrupole 1, quadrupole 2 and diagnostics 1 respectively. The diagnostics is located 1m after the wall. The beam pipe inside the wall is 78 mm diameter.

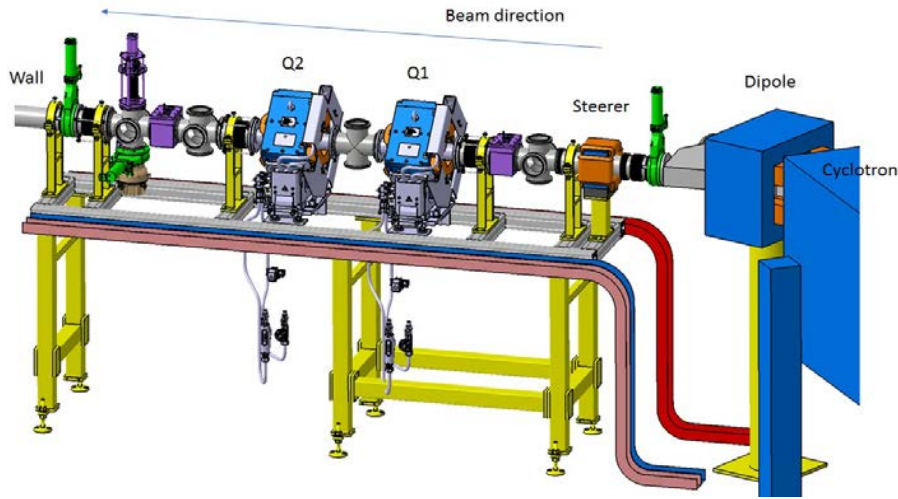

Figure 9: 3D view of the CBTL.Q1 and Q2 are quadrupoles.

For diagnostics and maintenance operations, a gap of 1 m is applied in the simulation between the wall and the entrance of the five-exit dipole. For the final design of the two experimental beamlines, the magnetic field map of the five-exit dipole has been obtained for the purpose of beam dynamics simulations.

Table 6: Main characteristics of the quadrupoles of the CBTL.

| Parameters | Values |
|---|---|
| Aperture radius (mm) | 39 |
| Max. gradient, G (T/m) | 7.34 |
| Magnetic length, $L_{eff}$ (mm) | 249.89 |
| Max. current (A) | 125 |
| dG/G | $8.71 \times 10^{-5}$ |
| Inductance (mH) | 12.8 |
| Water pressure (bars) | 5 |
| Weight (kg) | 160 |

Table 7: Main characteristics of the five-exit dipole.

| Parameters | 16.5 MeV (22°, 40°) | 25 MeV (22°, 40°) |
|---|---|---|
| B0 (T) | 0.256, 0.481 | 0.316, 0.594 |
| Radius (mm) | 2300, 1225 | 2300, 1225 |
| Gap (mm) | 60 | 60 |
| Current (A) | 62.6, 117.5 | 77.3, 145 |
| dBL/BL (V±20mm, H±34mm) | $\leq 5 \times 10^{-3}$ | $\leq 5 \times 10^{-3}$ |
| Magnetic length (mm) | 937.17, 909.57 | 937.15, 909.34 |
| Max. power (kW) | $\leq 1.4$ | $\leq 1.4$ |
| Cooling (l/mm) | 10 | 10 |
| Water pressure (bars) | $\leq 5$ | $\leq 5$ |
| Total weight (kg) | 4000 | 4000 |

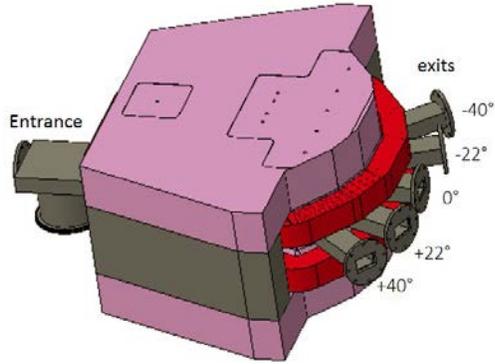

Figure 10: 3D view of the five-exit dipole.

### 3.5. *Design of the experimental area beamlines*

The main optics elements and the diagnostics of the CYRCé beamlines (CBTL, PRECy, and CMS sections) are shown in the figure below (Fig.11). The switching dipole has five exits, two of which already connected to the beam line sections of PRECy (-22°) and CMS (+22°). The former uses diffusers and collimators for defining the final beam properties, whereas the latter contains also two active elements, i.e. magnetic quadrupoles.

#### 3.5.1. PRECy beamline

The angle of deflection of the PRECy beamline is -22°. Considering the measured beam emittance, the desired lossless beam transport can be achieved by various combinations of quadrupole settings in the vault section. One can achieve smaller envelopes at the expense of a larger dispersion downstream the switching dipole and vice-versa. The following example is for I_QA = 51.0 A (corresponding to a field gradient of 3.00 T/m) and I_QB = 45.0 A (2.64 T/m) (Fig.12).

A FWHM beam size of 13 mm (horizontal) and 8 mm (vertical) has been measured and then estimated at the exit of the dipole by TraceWin multi-particle simulations. In order to deposit homogeneously the energy of the beam onto the samples, a diffuser is used after the steerer located at the exit of the dipole.

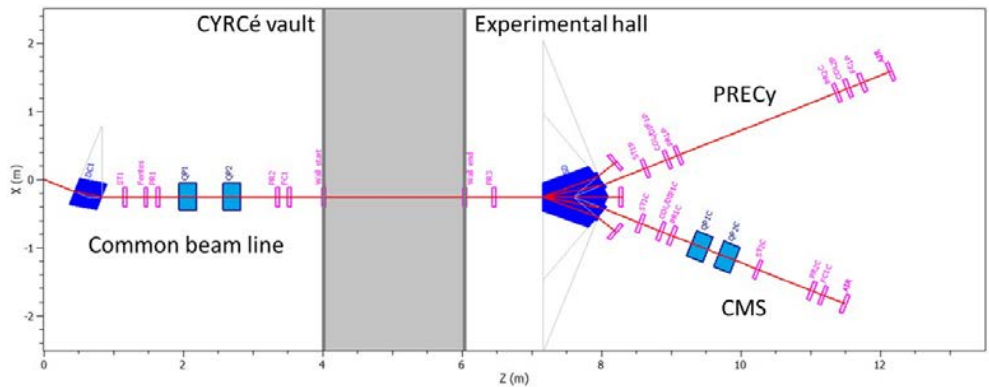

Figure 11: Beam line lattice containing main elements used in the simulations.

Simulations showed that a total length of 3 m should be sufficient for the beam to meet the requirements of the experimental station.

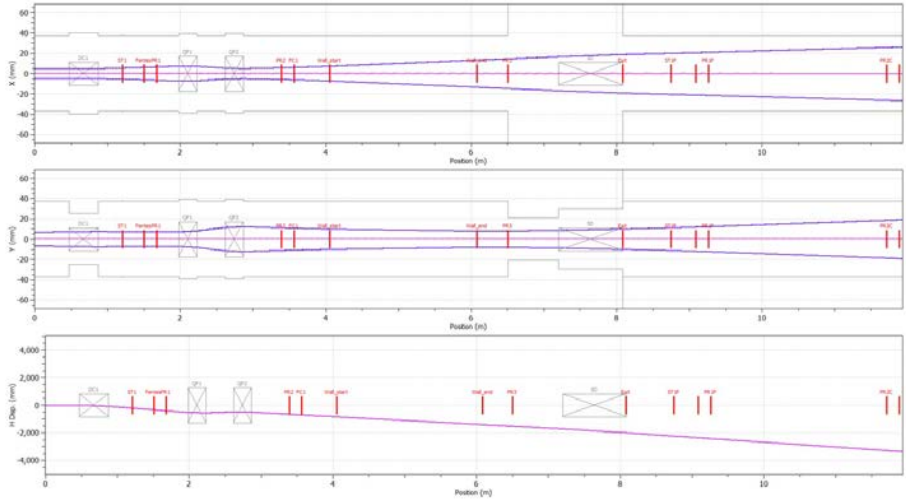

Figure 12: Beam optics of the PRECy beamline. Top two panels are for horizontal and vertical planes respectfully. The envelope lines correspond to 3$\sigma$ RMS. The lower panel shows the dispersion through the beam line.

Fig.13 shows the 3D view of the PRECy beamline. Profilers after the steerer and before the station measure the shape of the beam. The line is composed of a pumping system, valves, one XY steerer, one multi-position diffuser, collimators and one Faraday cup. The whole length of the beamline is 4906 mm (from the deviation point in the switching dipole).

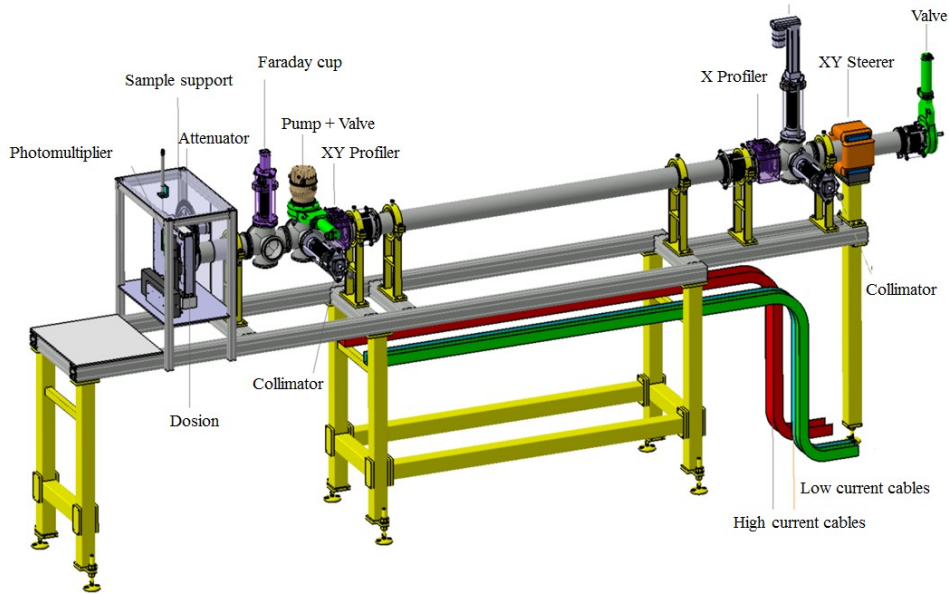

Figure 13: 3D view of the PRECy beamline.

*3.5.2. CMS beamline*

Similarly to the requirements for PRECy, for the CMS section the dispersion of the line has been considered while designing the CMS beamline. According to the simulations, the use of a pair of quadrupoles placed after a diffuser system is mandatory in order to meet the desired beam requirements at the CMS test station. Indeed, using these optical devices allows reaching a beam spot of 2.8 mm (RMS) in horizontal and 1.8 mm (RMS) in the vertical plane respectfully. The example shown in (Fig.14) is obtained for CMS quadrupoles settings at 3.23 T/m (55.04 A) and 3.93 T/m (66.89 A). Further reduction of the beam size can be achieved by the use of the collimators.

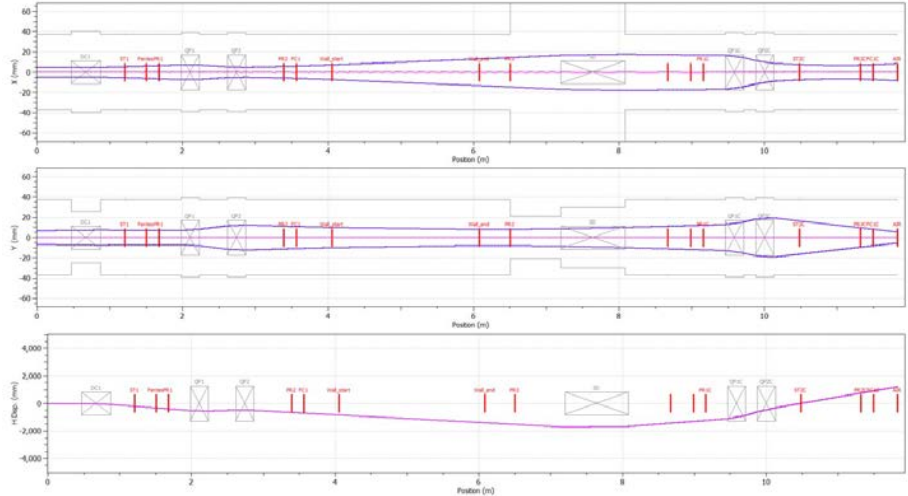

Figure 14: Beam optics of the CMS beamline. Top two panels are for horizontal and vertical planes respectfully. The envelope lines correspond to $3\sigma$ RMS. The lower panel shows the dispersion through the beam line.

The quadrupoles, manufactured by SigmaPhi, have the same characteristics as the ones used for the CBTL. As the use of quadrupoles may cause unwanted steering in case of transverse beam offsets, a XY correction steerer is placed between the experimental station and the doublet (Fig.15). The CMS beamline contain pumps, a Faraday cup and two profilers and is 4194 mm long, from the deviation in the dipole to the experimental station.
Fig.16 shows the entire set up from the cyclotron to the two experimental areas, including the CBTL, the five-exit dipole, the PRECy and the CMS beamlines.

## 4. Commissioning of the beamlines

Once the control system of the beam lines had been built and operational, the first beam was transported to the experimental areas at energy of 25 MeV. A synoptics of the beamlines are shown in Fig.17. Beam tuning of the CBTL is performed via a graphical user interface. The aim was to check that all device involved were properly set. Vertical slits protect the meander plates against direct beam losses. Scrapers and slit positions are set to remove the halo of the proton beam. The beamline is first tuned with the full beam current on the Faraday cup.

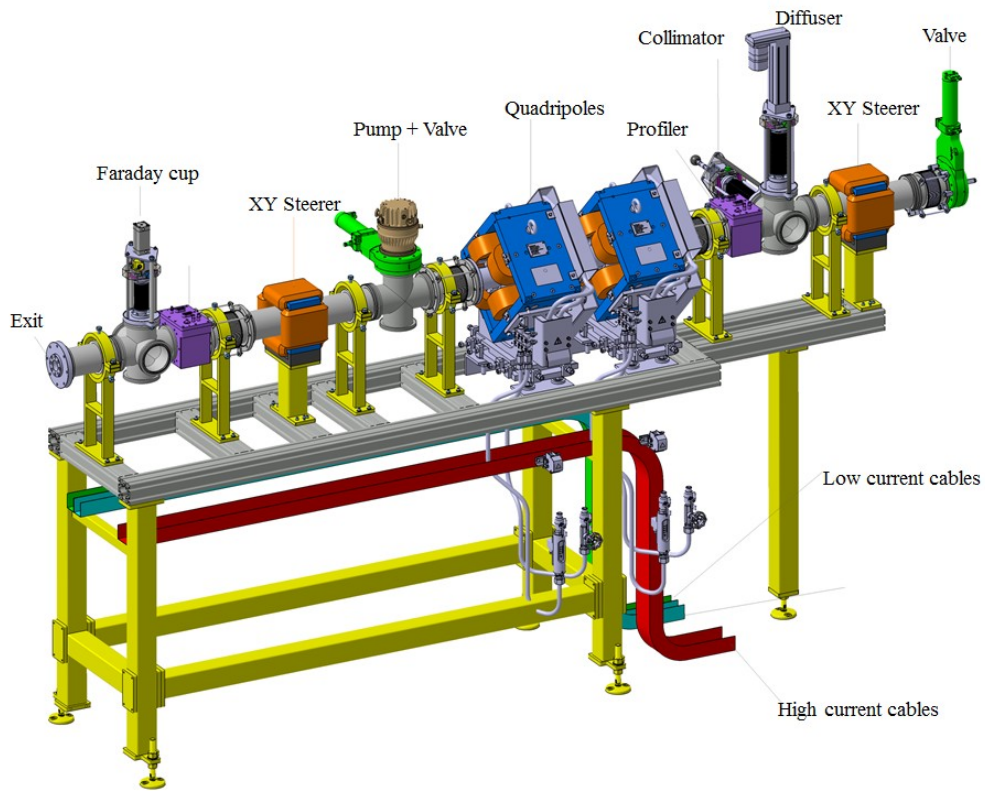

Figure 15: 3D view of the CMS beamline.

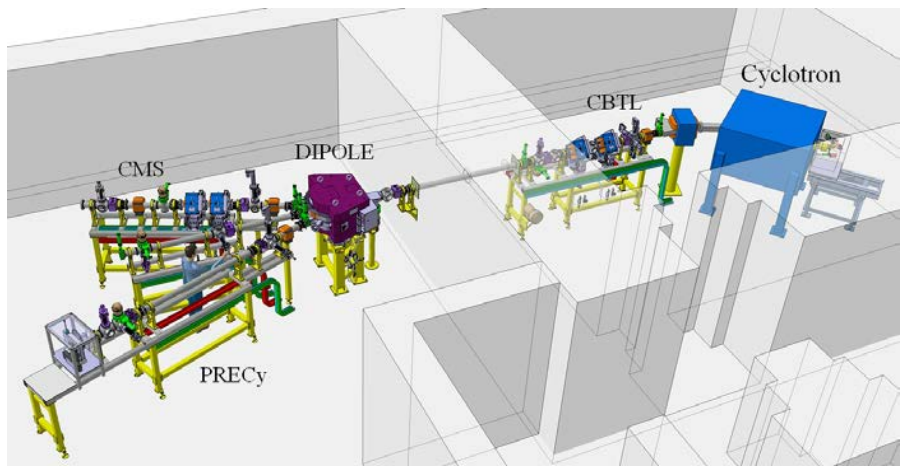

Figure 16: 3D view of overall experimental area.

A fast single wire scanning profiler with adjustable rotation, entirely designed and built at IPHC, allows measuring the beam profile after it passes the wall and before it enters the five-exit dipole (Fig.18). Indeed, this location is the main strategic part of the whole facility.

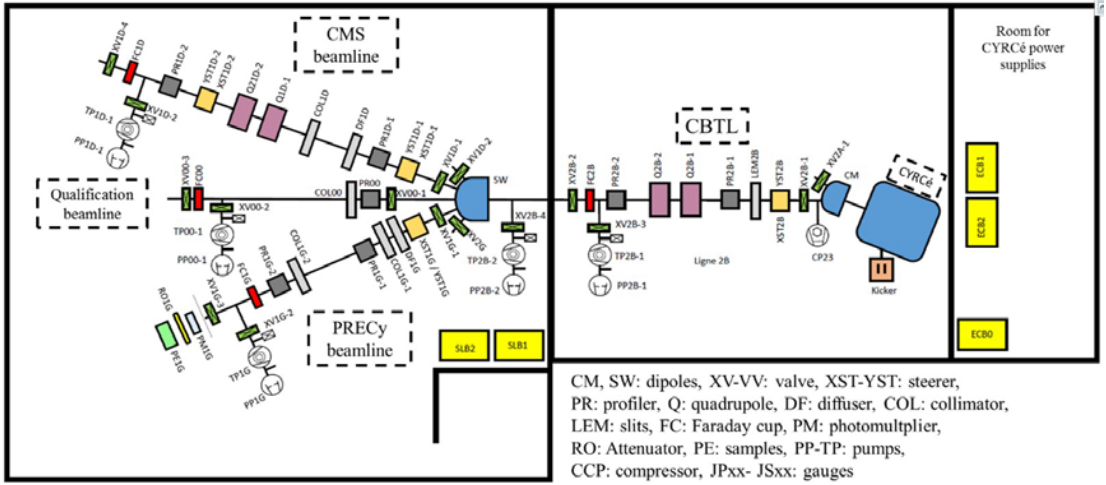

Figure 17: Synoptics of the CBTL, CMS and PRECy beamlines.

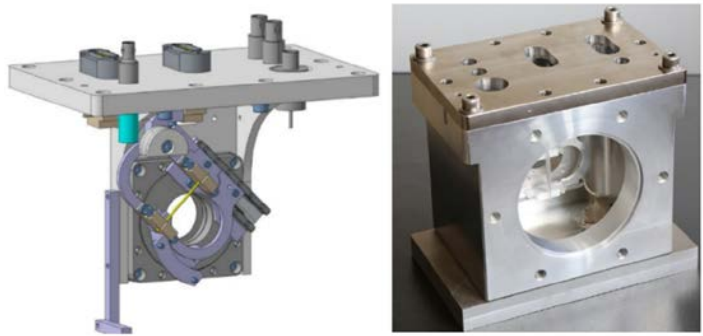

Figure 18: 3D view (left) and picture (right) of the angular single-wire scanner.

This profiler consists of two piezoelectric displacement axes: translation and rotation. Table 8 presents the characteristics of the device. The wire, made of tungsten, has rectangular cross-section with a width of 0.5 mm and is fixed to a Teflon support. The beam deposits its charge on the wire and the current is measured down to pA levels using I/V cards.

Measurements are done first when all quadrupoles of the CBTL are switched off then with current to adjust the beam and meet the requirements. Fig.19 shows the location of the different FCs and profilers along the CBTL.

Table 8: Characteristics of the scanning wire.

| Parameters | Values |
|---|---|
| Travel (mm) | ±20.05 (41) |
| Step width (mm) | 1-1.500 |
| Scan range (µm) | ≥1.5 |
| Scan resolution (nm) | ≤1 |
| Velocity (mm/s) | ≥20 |
| Max. working frequency (kHz) | 18.5 |

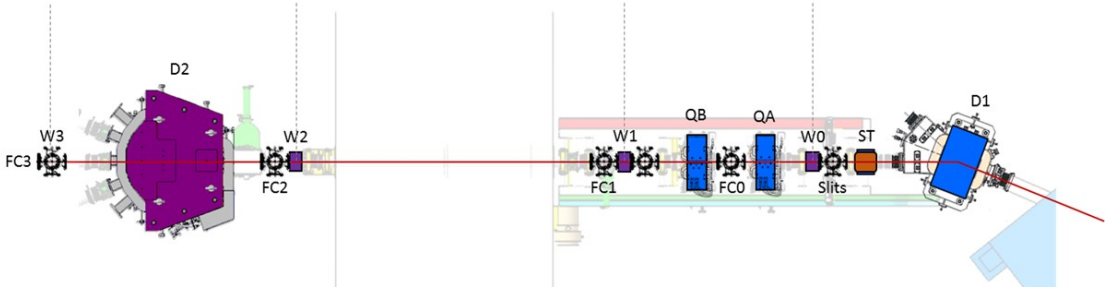

Figure 19: Simplified synoptics of the CBTL. Profilers: W0, W1, W2 and W3. Faraday cups: FC0, FC1, FC2 and FC3. Quadrupoles: QA and QB. Dipoles: D1 and D2. Slits are located between the steerer (ST) and the profiler W0.

The horizontal and vertical profiles of the beam are measured at several locations along the beam line: at the middle and at the end of the CBTL line and after the wall (Fig.20).

Fig.21 presents the transmission of the beam through the CBTL at different Faraday cup locations. Following the size of the aperture of the slits, no losses were observed along the line from FC0 to FC3.

### 4.1. PRECy beamline

To transport the protons of 25 MeV, up to the end of the PRECy beamline, the beam is first optimized in the CBTL using different profilers. Currents of 51 A and 45 A in the quadrupoles allows obtaining beam profiles quite well centered and with a Gaussian shape (Fig.22, left). Once the CBTL optimized, the protons are directed to the PRECy experimental section.

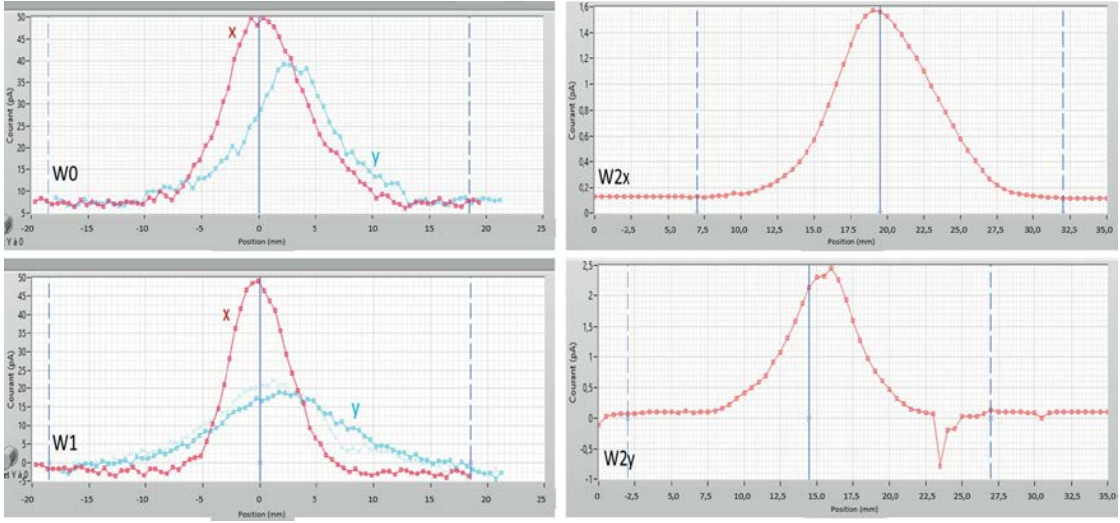

Figure 20: Tuning of the CBTL using the profilers, W0 (top, left) and W1 (bottom, left). The current in the quadrupoles QA and QB are 45 A and the slits are open at +/- 10mm. FWHMs of the beam are measured at 7.8 mm in x and 6.7 mm in y at W0 whereas they are 5.9 in x and 12.5mm in y at W1. Profile of the beam at the W2 location in the horizontal (top, right) and vertical (bottom, right) planes. FWHM are x 7.75 mm and y 5.3 mm.

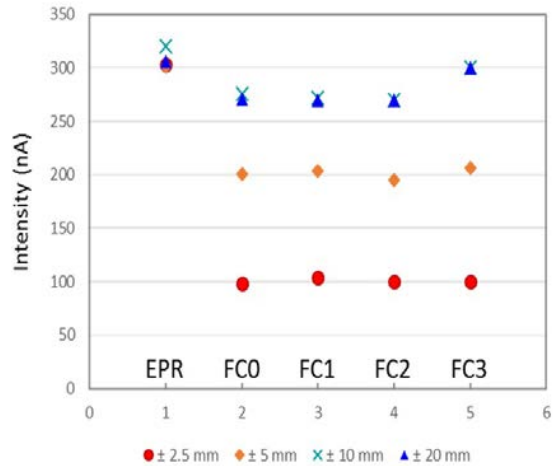

Figure 21: Beam transmission through the CBTL versus the aperture of the slits.

A current of 72 A in the five-exit dipole allows deflecting the beam from -22° to the right beamline. Beam profiles along the PRECy beamline (Fig.22) show a good agreement with the simulations. Beam dimensions measured meet the requirements stated at the beginning of the project. Indeed, minimum size of beam is measured at 7 mm diameter that is below and better than the 24 mm diameter required. Before reaching the experimental location, the proton will be diffused in order to have a homogeneous distribution.

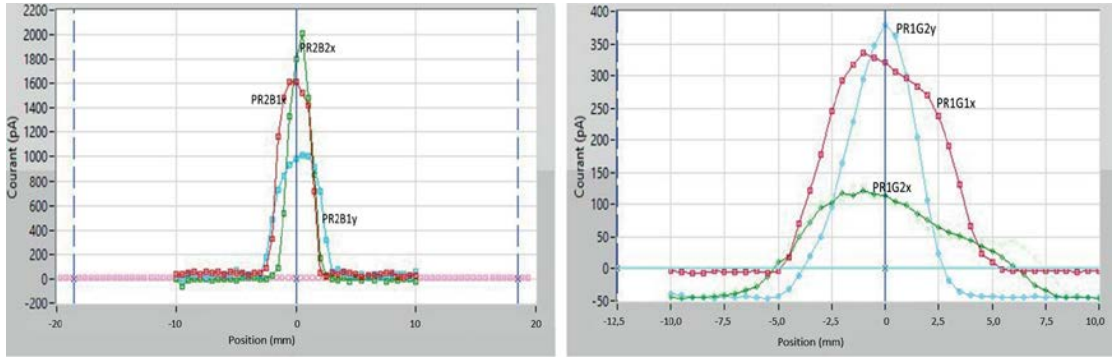

Figure 22: Horizontal and vertical profiles of the beam at two different locations along the CBTL (left) and at two different locations along the PRECy beamline (right). PR2B1 refers to W0 seen previously and PR2B2 refers to W1. The profiler PR1G1 is located just after the five-exit dipole whereas PR1G2 is located at the end of the beamline (See Fig. 17). Proton beam at 25 MeV.

### 4.2. CMS beamline

To transport the protons of 25 MeV, up to the end of the CMS beamline, the beam is first optimized in the CBTL using the same parameters as used while commissioning the PRECy line. Currents of 51 A and 45 A in the quadrupoles allows obtaining beam profiles centered and with a Gaussian shape (Fig.23, left).

Once the CBTL optimized, the protons are directed to the CMS experimental section. An inverted current of 72 A in the five-exit dipole allows deflecting the beam from +22°. An optimal focusing is achieved when setting currents of 65 A and 70 A in the magnets of the quadrupoles of the CMS beamline.

Plots on Fig.23 (right), show the transverse profiles of the beam at the beginning of the CMS beamline and after the doublet of quadrupoles composing it. Although a double structure of the beam is suggested in the horizontal plane by the profile measurement at PR1D2x, beam dimensions meet the requirements and show a good agreement with the simulations. Indeed, a minimum of 4 mm and 10 mm beam diameter in the horizontal and vertical planes respectively are measured which is below the 30 mm diameter required for the experiment.

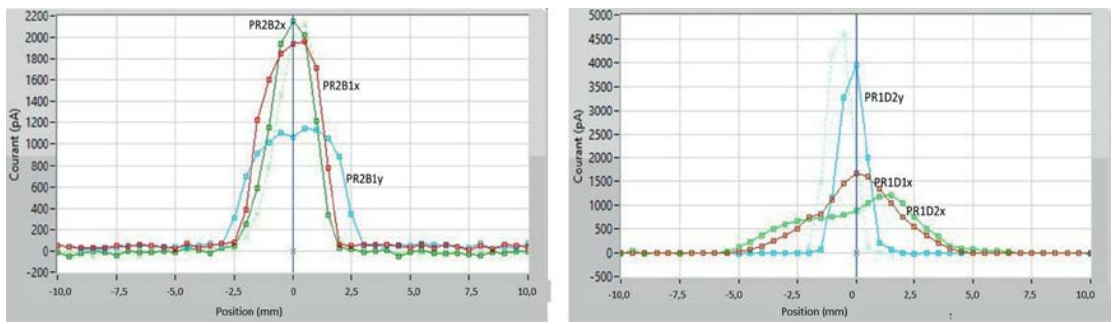

Figure 23: Horizontal and vertical profiles of the beam at two different locations along the CBTL (left) and at two different locations along the CMS beamline (right). PR2B1 refers to W0 seen previously and PR2B2 refers to W1. The profiler PR1D1 is located just after the five-exit dipole whereas PR1D2 is located at the end of the beamline (See Fig. 17). Proton beam at 25 MeV.

A collimating system will cut the double structure to ensure a good homogeneity of the beam before starting the experiment. Additionally, the double structure was found to be highly dependent on the cyclotron extraction parameters. A clean, single-peak distribution can be achieved easily by adjusting the extraction foil and main magnet settings at the expense of a lower beam intensity in the CBTL. The latter though is not a problem due to the possibility to increase proton current at the injection without introducing unwanted beam structures.

## 5. Conclusions

The first two CYRCé extension beamlines have been designed, developed and operational since January 2020. One beamline is dedicated to radiobiological experiments (PRECy project) and the other to the tests of silicon modules (CMS project) (Fig.24). The new beamlines started with the development of a beamline (CBTL) in the vault, transporting the protons from the cyclotron to the experimental area (Fig.25). The measurement of the transverse emittances of the protons extracted from the accelerator was mandatory to design the CBTL. Several techniques have been employed to obtain these parameters. The main elements of the CBTL are two quadrupoles, a steerer and few profilers to manage and ensure the quality of the beam before it reaches a five-exit dipole that will deflect the particles to the dedicated experimental stations. The PRECy and the CMS beamlines have been each designed according to defined specifications. The CMS line section is composed of two steerers and a doublet of quadrupoles whereas the PRECy beamline is composed of steerers and diffuser to achieve a distribution the most homogeneous possible at the experiment station. The commissioning of the entire facility shows a good agreement with the beam dynamics simulations and meets the requirements stated at the beginning of the project.

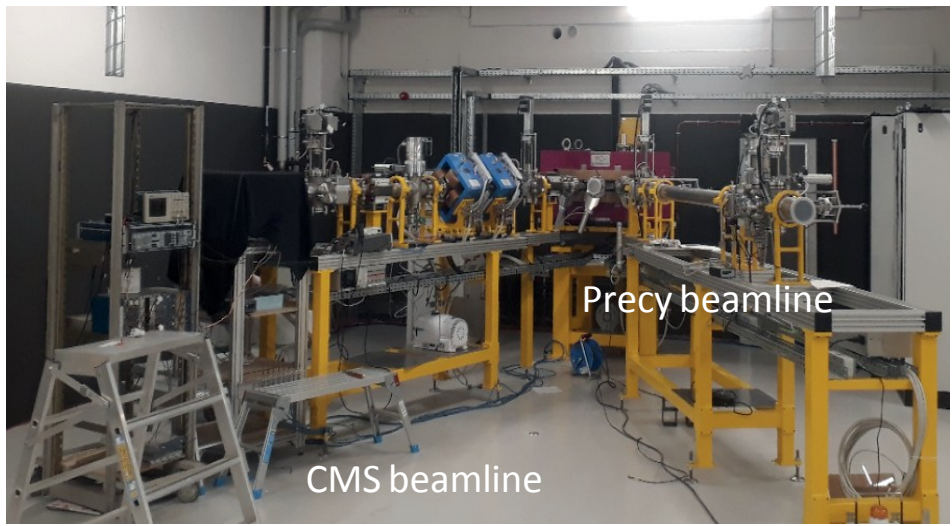

Figure 24: Picture of the CMS and PRECy beamlines

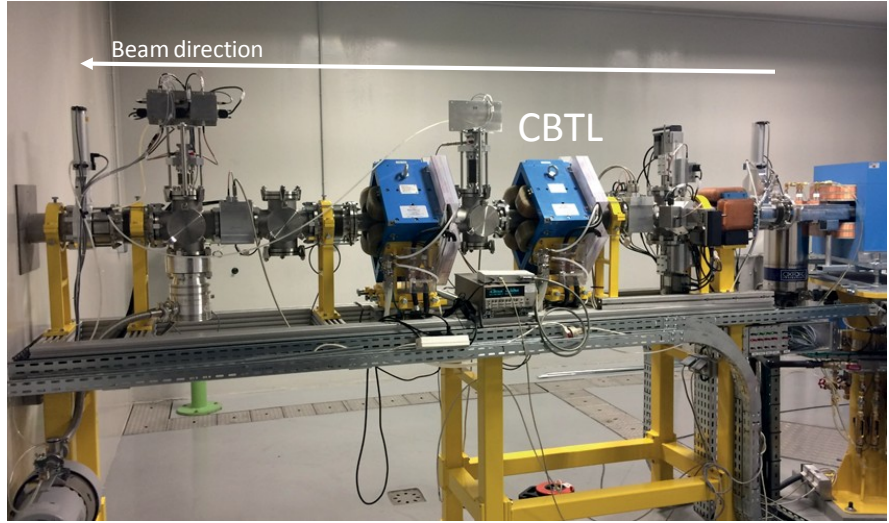

Figure 25: Picture of the CBTL

## 6. Acknowledgments

This achievement is the result of a fruitful collaboration between several scientific and technical teams from IPHC. The authors would like to thank the following teams: *Hadrontherapie*, *Plateforme CYRCé*, *Service Mécanique*, *Instrumentation des Accélérateurs*, *Systèmes de Mesure et d'Acquisition*, *Service de Radioprotection*, *Service Informatique* and the CMS team. The PRECy project is supported by the *Contrat de Projet Etat-Région* (CPER) number 918-15 C1, *Région Grand-Est*, 2015-2020, and *Eurométropole Strasbourg* (EMS), CNRS, France. The CMS beamline is funded by IN2P3/CNRS, France.